\RequirePackage{xcolor}
\documentclass[journal,twoside,web]{ieeecolor}
\usepackage{tmi}

\usepackage{etoolbox}
\makeatletter
\@ifundefined{color@begingroup}%
  {\let\color@begingroup\relax
   \let\color@endgroup\relax}{}%
\def\fix@ieeecolor@hbox#1{%
  \hbox{\color@begingroup#1\color@endgroup}}
\patchcmd\@makecaption{\hbox}{\fix@ieeecolor@hbox}{}{\FAILED}
\patchcmd\@makecaption{\hbox}{\fix@ieeecolor@hbox}{}{\FAILED}

\usepackage{cite}
\usepackage{amsmath,amssymb,amsfonts}
\usepackage{algorithmic}
\usepackage{graphicx}
\usepackage{dblfloatfix}
\usepackage{textcomp}
\usepackage{lipsum}
\usepackage{soul}
\usepackage{booktabs}
\usepackage{tikz}
\definecolor{darkred}{rgb}{.0,.0,.0} 
\definecolor{darkgreen}{rgb}{.0,.0,.0}

\definecolor{xkcdgreen}{rgb}{.08,.69,.10}
\definecolor{xkcdorange}{rgb}{.98,.45,.0}
\definecolor{xkcdred}{rgb}{.9,.0,.0}
\definecolor{xkcdpurple}{rgb}{.5,.12,.61}

\newcommand{\PHIF}{\Phi_{\mathcal{F}}}
\newcommand{\PHIB}{\Phi_{\mathcal{B}}}

\newcommand{\minus}{\scalebox{0.75}[1.0]{$-$}}

\def\BibTeX{{\rm B\kern-.05em{\sc i\kern-.025em b}\kern-.08em
    T\kern-.1667em\lower.7ex\hbox{E}\kern-.125emX}}
\newcommand\copyrighttext{%
  \footnotesize \textcopyright 2023 IEEE. Personal use of this material is permitted. Permission from IEEE must be obtained for all other uses, including republication/redistribution.}
\newcommand\copyrightnotice{%
\begin{tikzpicture}[remember picture,overlay]
\node[anchor=south,yshift=0.5cm] at (current page.south) {\fbox{\parbox{\dimexpr\textwidth-\fboxsep-\fboxrule\relax}{\copyrighttext}}};
\end{tikzpicture}%
}

\usepackage{hyperref}
\begin{document}

\title{Robust deformable image registration using cycle-consistent implicit representations}

\author{Louis D. {van Harten}, Jaap {Stoker}\ and Ivana I{\v s}gum
\thanks{Manuscript received December 23rd, 2022; revised June 9th, 2023; accepted September 25th, 2023. This is the accepted version of the manuscript. \textit{(Corresponding author: Louis D. van~Harten)}}
\thanks{Louis. D. van Harten and Ivana I{\v s}gum are with the Department of Biomedical Engineering and Physics, Amsterdam UMC location University of Amsterdam, the Netherlands and with the Informatics Institute, University of Amsterdam, the Netherlands (e-mail: l.d.vanharten@amsterdamumc.nl; i.isgum@amsterdamumc.nl). }
\thanks{Jaap Stoker and Ivana I{\v s}gum are with the Department of Radiology and Nuclear Medicine, Amsterdam UMC location University of Amsterdam, the Netherlands. Jaap Stoker is also with Amsterdam Gastroenterology Endocrinology Metabolism, Amsterdam, the Netherlands and with Cancer Center Amsterdam, Amsterdam, the Netherlands (e-mail: j.stoker@amsterdamumc.nl).
}
}

\maketitle
\copyrightnotice

\begin{abstract}
Recent works in medical image registration have proposed the use of Implicit Neural Representations, demonstrating performance that rivals state-of-the-art learning-based methods. However, these implicit representations need to be optimized for each new image pair, which is a stochastic process that may fail to converge to a global minimum. 
To improve robustness, we propose a deformable registration method using pairs of cycle-consistent Implicit Neural Representations: each implicit representation is linked to a second implicit representation that estimates the opposite transformation, causing each network to act as a regularizer for its paired opposite. During inference, we generate multiple deformation estimates by numerically inverting the paired backward transformation and evaluating the consensus of the optimized pair. This consensus improves registration accuracy over using a single representation and results in a robust uncertainty metric that can be used for automatic quality control.
We evaluate our method with a 4D lung CT dataset. The proposed cycle-consistent optimization method reduces the optimization failure rate from 2.4\% to 0.0\% compared to the current state-of-the-art. The proposed inference method improves landmark accuracy by 4.5\% and the proposed uncertainty metric detects all instances where the registration method fails to converge to a correct solution. We verify the generalizability of these results to other data using a centerline propagation task in abdominal 4D MRI, where our method achieves a 46\% improvement in propagation consistency compared with single-INR registration and demonstrates a strong correlation between the proposed uncertainty metric and registration accuracy.
\end{abstract}


\section{Introduction}
\label{sec:introduction}
\IEEEPARstart{I}n image registration, a spatial mapping between different images is found by placing them in a shared coordinate space while maximizing local correspondences of image content. Registration is an important part of many image analysis pipelines; it is used for aligning multi-modal data for joint analysis~\cite{wells1996multi}, for motion analysis in time-series data~\cite{wang2011cardiac}, for disease development tracking in follow-up imaging~\cite{castadot2010assessment} and for various other tasks. 
Due to the high potential impact of mistakes in clinical practice, output of automatic methods is generally subjected to human quality control.
However, such quality control may be unfeasible in systems where a real-time method output relies on an accurate registration (e.g. real-time analysis in ultrasound imaging~\cite{narang2021utility}), or when analysis requires a large number of image registrations (e.g. motion analysis in time-series~\cite{de2018evaluation}). In such cases, robustness of a registration method is especially important. The lack of this property has been one of the main hurdles for introducing modern AI systems in clinical practice~\cite{he2019practical}.

In recent years, numerous deep-learning based methods have been proposed for pairwise deformable image registration\textcolor{darkred}{~\cite{de2019deep,balakrishnan2019voxelmorph,haskins2020dldirsurvey,hering2022learn2reg}}. The majority of these methods are grid-based: a CNN is used to extract features from both images and predict a rasterized vector field that maps one image to the coordinate space of the other. A downside of such an approach is that the trained CNNs tend to be highly sensitive to changes in the rasterized features in either image domain, resulting in systematic fragility to out-of-distribution data. \textcolor{darkred}{To mitigate the impact of this problem, recent works have employed test-time optimization to fine-tune registration CNNs~\cite{zhu2021test}. However, this approach requires optimizing a large network at test-time, which eliminates the efficiency benefits of learning-based methods over optimization-based ones.}

\textcolor{darkred}{By formulating image registration as a continuous function operating on coordinates,} a deformation field can be optimized as an Implicit Neural Representation (INR): a neural network, applied as a function approximator to map continuous coordinates to their respective deformation vectors. INRs have shown great promise in representing various spatial signals, such as shape descriptors of objects~\cite{park2019deepsdf} and natural image scenes~\cite{liu2020neural,mildenhall2021nerf}. 
Recently, Wolterink et al. proposed a deformable registration method for medical images using INRs~\cite{wolterink2021implicit}. This method uses sinusoidal representation networks (\textsc{SIREN}s)\cite{sitzmann2020implicit}, which allow networks to capture higher-order spatial derivatives. This enables analytic computation of second-order regularization terms and makes it easier to represent small details in the vector fields compared to ReLU-based INRs\textcolor{darkred}{, which have a tendency to under-represent high-frequency signal components~\cite{tancik2020fourier}}. This registration method has been shown to perform on par with state-of-the-art learning-based methods \textcolor{darkred}{in 4D lung CT registration.}
\textcolor{darkred}{A similar approach has been extended with meta-learning for efficient registration of 3D ultrasound images~\cite{baum2022meta}.}

\begin{figure}[!t]
\centering
\includegraphics[width=8.5cm]{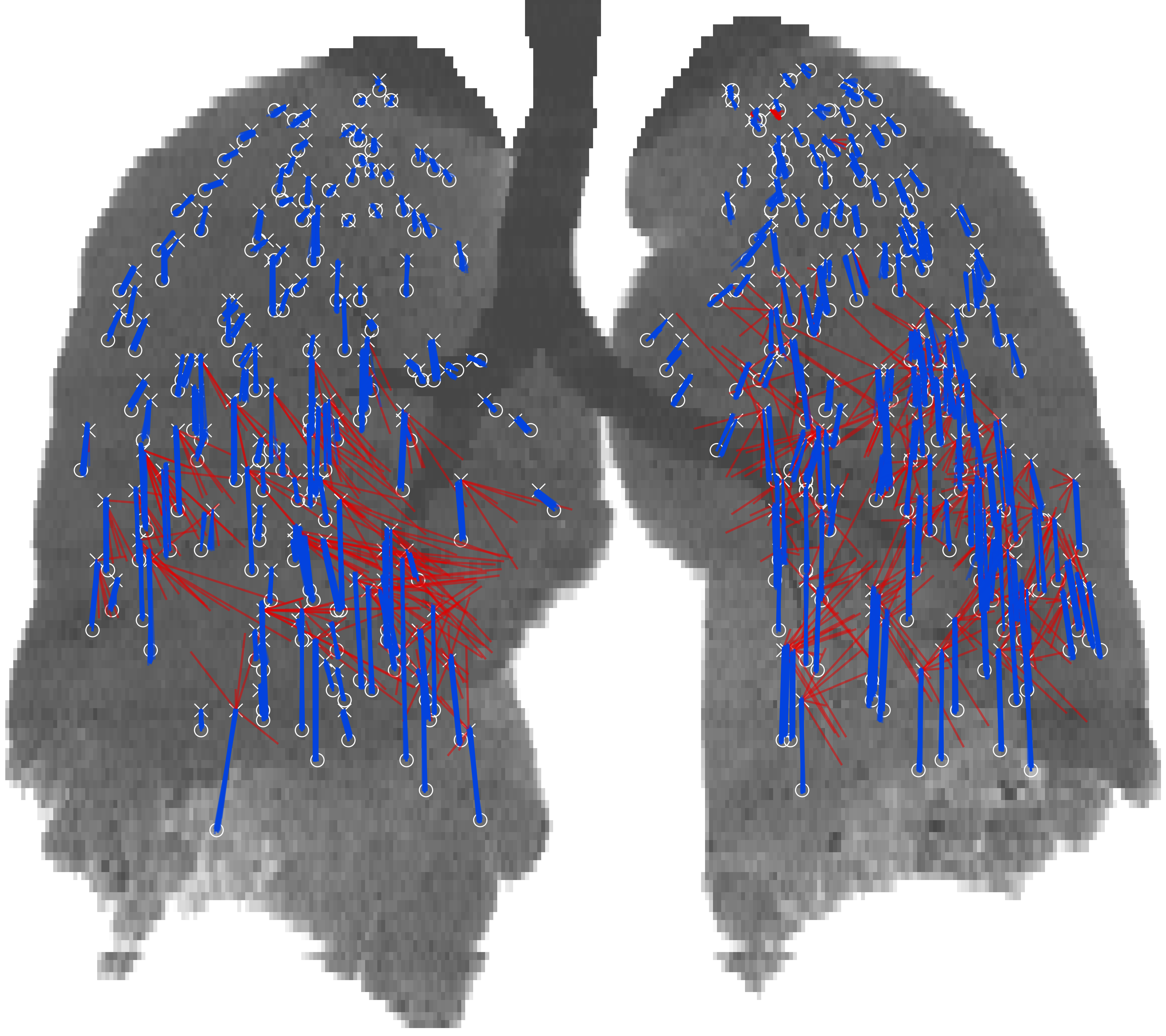}
\caption{Results of 100 deformable registrations obtained from implicit representations optimized \textit{without} cycle-consistent regularization, initialised using different random seeds. Projected ground-truth landmarks indicated by white markers, projections of registration results for each landmark and for each \textcolor{darkred}{run} shown as lines. Results with errors over five millimeters are coloured red, results with smaller errors are coloured blue. Most lines are invisible due to near-perfect overlap, but a few \textcolor{darkred}{runs} diverge.}
\label{fig:dirlab_example_cc0}
\end{figure}

While the results attained by the abovementioned method are promising, the method is not guaranteed to converge to a correct solution. Depending on the images at hand and on the network initialization, the optimization of the implicit registration function may collapse in a local minimum.
This is demonstrated in Fig. \ref{fig:dirlab_example_cc0}, which shows 
multiple registration results 
obtained using different random seeds: most \textcolor{darkred}{runs} yield a near-identical solution, but some fail to converge to a correct result. \textcolor{darkred}{This failure rate can be considered a measure of robustness. The observed failures are} especially problematic in autonomous settings where a pipeline of image analysis tools would depend on the correctness of \textcolor{darkred}{the} registration.

In this work, we employ cycle-consistency to develop a robust optimization method for registration using INRs. \textcolor{darkred}{This approach was inspired by cycle-consistent image registration~as proposed in Christensen and Johnson~\cite{christensen2001consistent}, which has been~shown to result in more anatomically plausible and reproducible deformation fields. Similar effects have been observed when cycle-consistency was applied to learning-based registration methods~\cite{kim2019unsupervised,kim2021cyclemorph}. Given these findings, we investigated whether using a cycle-consistent approach can make the INR optimization process more robust. Hence, we propose} a set of coupled, cycle-consistent INRs where one INR is optimized to estimate the forward transformation and the other is optimized to estimate the backward transformation. 
The backward transformation yields an estimate of the original input coordinates when applied to the output of the forward transformation, resulting in a cycle-error in the original coordinate space. The~two transformations can be linked by an optimization term minimizing this cycle-error, which allows the two INRs to regularize each other. We aim to stabilize the optimization procedure
\textcolor{darkred}{and reduce the failure rate}
by leveraging this cycle-consistent regularization~effect.

Taking advantage of the multiple-differentiable property of \textsc{SIREN}s, we use Taylor expansions to derive a highly accurate local approximation of the inverse of the backward transformation during inference. This yields a second estimate for the forward transformation, which is used to improve registration accuracy compared to single-INR inference. Additionally, the disagreement between these estimates yields a dense, highly interpretable uncertainty metric that can be used for automatic quality control.

We compare our cycle-consistent method to single-INR registration and we demonstrate improvement both in terms of registration accuracy and robustness. Furthermore, we evaluate the resulting uncertainty metric by correlating it to the errors in our results. To demonstrate that our results generalize with respect to data, we evaluate our method on both the DIR-LAB 4D CT lung dataset~\cite{castillo2009framework}, as well as on an intestinal motility dataset containing 4D MRI sequences of the abdomen~\cite{de2019detecting}. 
\textcolor{darkred}{The implementation of our method is publicly available}\footnote{\url{https://github.com/louisvh/cycle_consistent_INR}}.

\section{Data}\label{sec:data}
Two datasets are used in this work: the DIR-Lab 4D lung CT dataset\cite{castillo2009framework} and a dataset of abdominal 4D MRI scans~\cite{de2019detecting}. These datasets were selected to demonstrate the flexibility of the presented method, as the motion properties in both datasets are considerably different. The lung CT dataset yields vector fields with a large low-frequency component due to the uniform nature of the compression and expansion of the lungs. Conversely, the abdominal MRI dataset yields vector fields with large high-frequency components, as the dominant motion in these sequences is local squeezing motion of the intestines.

\subsection{4D lung CT}\label{sec:data_4DCT}
The publicly available DIR-Lab dataset~\cite{castillo2009framework} contains 10 axial 4D lung CTs with 10 time points across a shallow breathing cycle. In-plane resolution ranges from 0.97 mm to 1.16 mm isotropic, 0.97 mm being the most common. Slice thickness is 2.5 mm in all cases. At maximum inspiration and expiration, 300 manually annotated lung landmarks are available for each scan, serving as ground-truth for evaluating registration methods between the two time points.

\begin{figure*}[!t]
\centering
\includegraphics[width=0.64\linewidth]{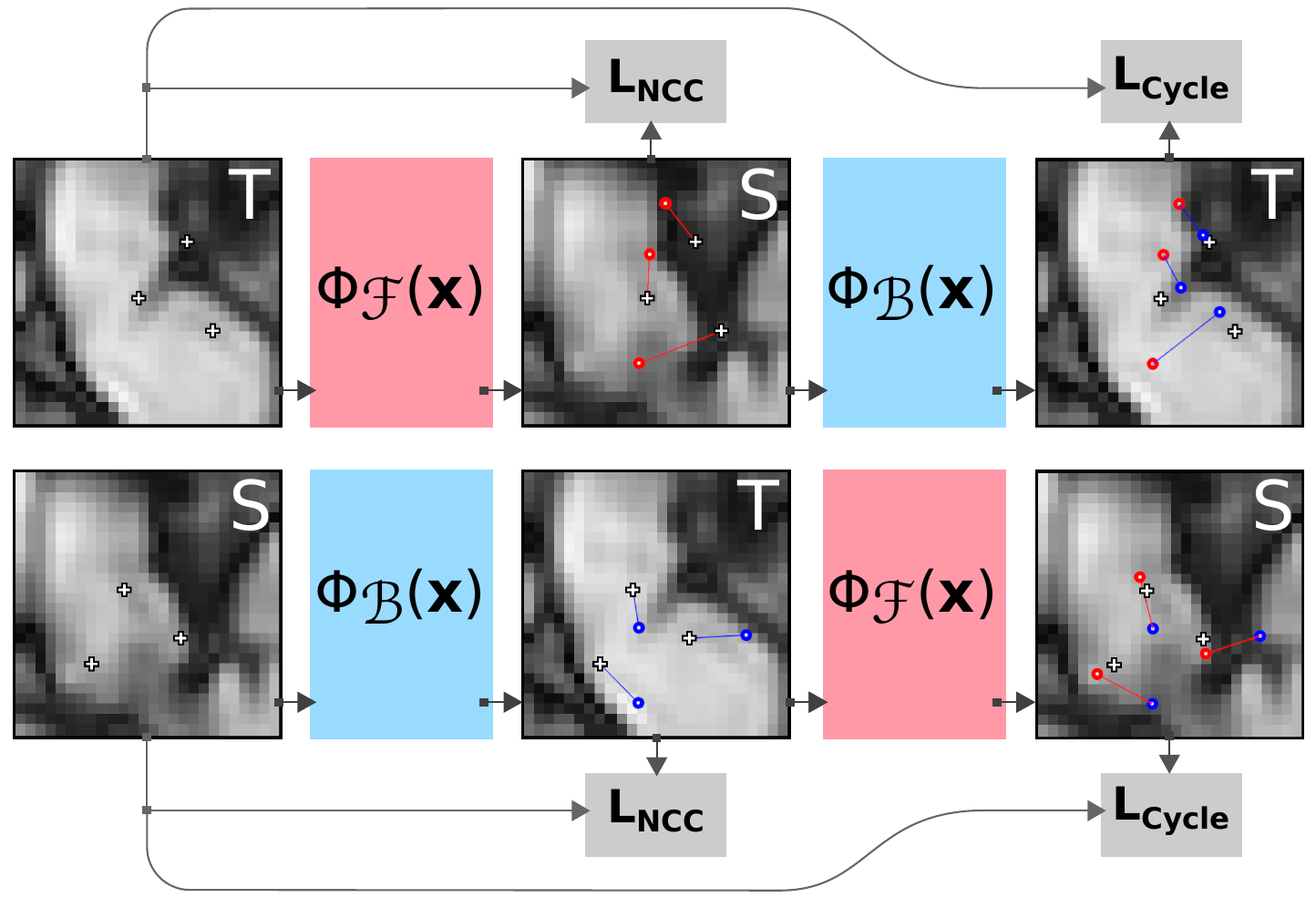}
\caption{A schematic overview of the optimization procedure for the cycle-consistent implicit neural representations. The forward representation ($\PHIF$) maps the deformation field from the target to the source image and the backwards representation ($\PHIB$) maps the field from the source to the target image. The NCC loss components operate on the mapped intensity values, whereas the cycle-consistency loss operates on the euclidean vectors from each sample $\vec{x}_{T}$ to $\PHIB(\PHIF(\vec{x}_{T}))$ and from $\vec{x}_{S}$ to $\PHIF(\PHIB(\vec{x}_{S}))$, respectively. Note that while the images in the figure show a 2D projection, the method operates in 3D space.}
\label{fig:cc_training}
\end{figure*}

\subsection{Abdominal 4D MRI}\label{sec:data_4DMR}
The abdominal 4D MRI dataset~\cite{de2019detecting} contains scans of 14 healthy volunteers, consisting of volumetric sequences acquired at 1.0 volume per second during a breath-hold. Each volume was acquired at a resolution of 2.5x2.5x2.5~mm and reconstructed to 1.4x1.4x2.5~mm with an FOV of 400x400x35~mm. As the outer two slices contain severe imaging artifacts in the majority of volumes, these slices were cropped from the images. Small intestine centerline segment annotations are available for one volume in each sequence, as detailed in~\cite{van2022untangling}. Each sequence contains at least fifteen volumes after the timepoint for which centerline annotations are available. Due to the difficulty of the annotation task, the centerline annotations are not perfectly centered in the intestines. As centerline annotations sometimes extend outside of the imaged volume, these were cropped to the valid image domain.

\section{Methods}\label{sec:methods}
We present a method for deformable image registration using cycle-consistent sets of Implicit Neural Representations (INRs). By conditioning the optimization procedure of an INR on a simultaneously optimized backward function, we aim to reduce the instability of the system. To improve registration accuracy, we employ an estimate of the inverse of the backward function to derive a consensus for each forward result during inference. The magnitude of the difference of both estimates is used as an uncertainty metric.

\subsection{Cycle-consistent implicit representations}\label{sec:cc_optim}
Implicit Neural Representations are neural networks applied as universal function approximators. Rather than operating on pixel values, these networks operate on continuous image coordinates. The networks are not exposed to the pixel values directly: the image information only interacts with the network weights through backpropagation of gradients from a loss function that relates coordinates to pixel values. 

We use implicit neural representations $\Phi(\bar{x}) = \bar{x}+u(\bar{x})$ to parameterize transformation functions between image domains.
We define forward transformation $\PHIF(\vec{x}) = \vec{x} + u_{\mathcal{F}}(\vec{x})$, mapping coordinates in the target domain to coordinates in the source domain, and backward transformation $\PHIB(\vec{x}) = \vec{x} + u_{\mathcal{B}}(\vec{x})$, mapping coordinates in the source domain to coordinates in the target domain, where the functions $u_{\mathcal{F}}$ and $u_{\mathcal{B}}$ map coordinates to their corresponding deformation vectors. These functions are parameterized as sinusoidal representation networks (\textsc{SIREN}s\cite{sitzmann2020implicit}), which are lightweight multi-layer perceptrons that use sinusoidal activation functions. $\PHIF$ and $\PHIB$ are optimized jointly, extending the joint estimation approach from\cite{christensen2001consistent} by computing cycle-consistency terms explicitly, rather than numerically inverting both functions at each optimization step.

The proposed optimization method is visualized in Fig.~\ref{fig:cc_training}. 
Our total optimization objective consists of six components:

\begin{equation}
\begin{split}
\textit{L}_{total} =\ &L^{data}_{\mathcal{F}} + \alpha L^{reg}_{\PHIF} + \beta L^{cycle}_{\mathcal{F}\rightarrow\mathcal{B}} \ +\\
&L^{data}_{\mathcal{B}} + \alpha L^{reg}_{\PHIB} + \beta L^{cycle}_{\mathcal{B}\rightarrow\mathcal{F}}
\end{split}
\label{eq:totalloss}
\end{equation}

where $\alpha$ and $\beta$ are weighing factors for the regularization components. 
The data losses $L^{data}_{\mathcal{F}}$ and $L^{data}_{\mathcal{B}}$ are defined as 

\begin{equation}
\begin{split}
L^{data}_{\mathcal{F}} = \frac{2}{bs}\ \sum_{i=1}^{bs/2}\ &\minus NCC(S[\vec{x}_i], T[\PHIF(\vec{x}_i)]), \\
L^{data}_{\mathcal{B}} = \frac{2}{bs} \sum_{i=bs/2}^{bs} &\minus NCC(T[\vec{x}_i], S[\PHIB(\vec{x}_i)]),
\end{split}
\label{eq:dataloss}
\end{equation}

\begin{figure*}[!t]
\centering
\includegraphics[width=0.95\linewidth]{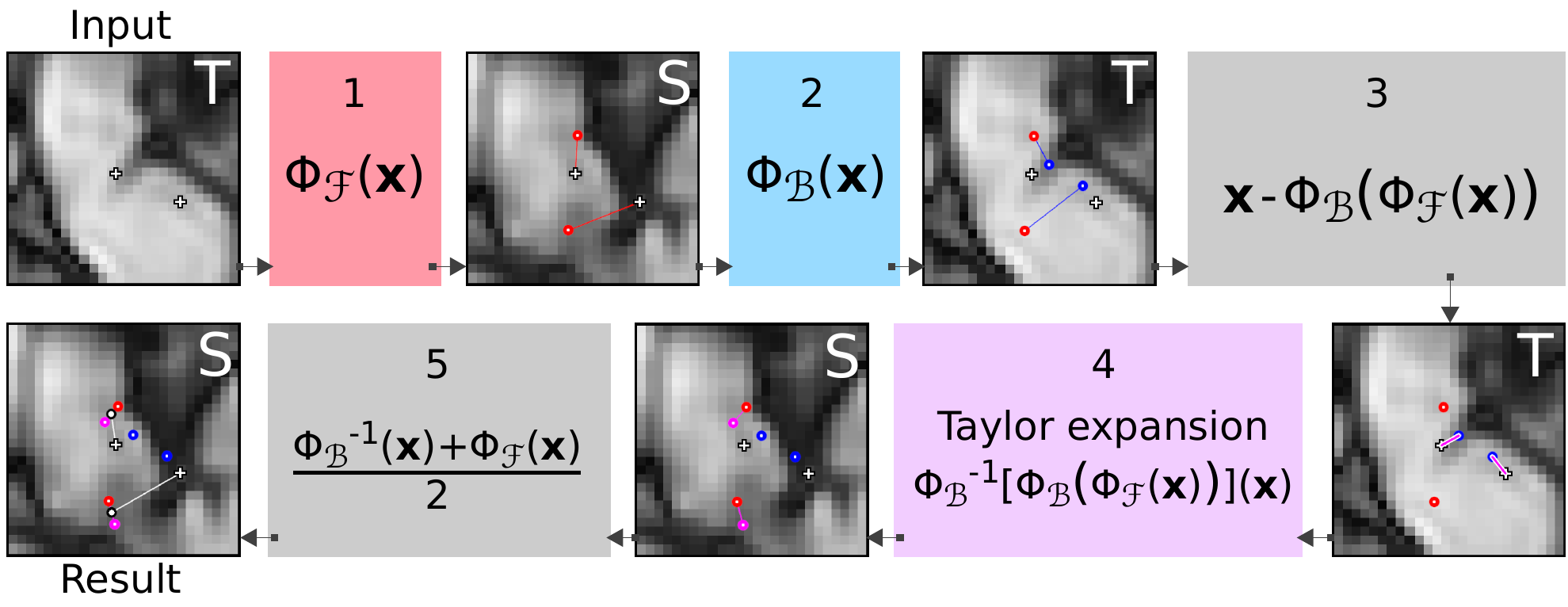}
\caption{An illustration of the cycle-consistent inference method for two arbitrary coordinates in the small bowel. (1) The forward INR $\PHIF$ is used to transform each coordinate $\vec{x}$ from the target domain to the source domain. (2) Backward INR $\PHIB$ is used to transform each coordinate $\PHIF(\vec{x})$ back to the target domain. (3) This yields cycle-error vector $\vec{x} \minus \PHIB(\PHIF(\vec{x}))$. (4) By Taylor expansion of $\PHIB^{\minus1}$ around each coordinate $\PHIB(\PHIF(\vec{x}))$, the cycle-error vector is projected from the target domain to the source domain, yielding the estimate $\PHIB^{\minus1}(\vec{x})$. (5) The midpoint of the vector between both estimates is taken as the final result. The vector norm is taken as a measure of uncertainty, analogous to an estimated confidence interval.}
\label{fig:cc_inference}
\end{figure*}

where $NCC$ is the normalized cross-correlation, $S$ is the source image, $T$ is the target image, $\vec{x}$ are sampled coordinates and $bs$ is the batch size. The first half of each batch is sampled from the valid domain of the source image, the second half is sampled from the valid domain of the target image. In the case of lung registration, these valid domains are defined as the lung masks in each timepoint. In the case of bowel registration, the valid domains are defined as the body mask, excluding the outermost slices.

As regularization term $L^{reg}_{\Phi}$ on each function, we employ a Jacobian determinant regularization. The Jacobian determinant of a function measures the local volumetric change in the vector field, indicated by its distance from $1.0$, yielding a simple regularization term:

\begin{equation}
    \textit{L}^{jac}[\Phi]= \frac{1}{bs} \sum_{i=1}^{bs} | 1 - \det \nabla \Phi[\bar{x_i}] |.
\label{eq:jacreg}
\end{equation}

However, this distance is not symmetric. While the distance for a factor $n$ growth is $n\minus1$, this distance for a factor $n$ shrinkage is merely $1\minus\frac{1}{n}$, yielding an asymmetric objective. Instead, we define a symmetric Jacobian determinant regularization

\begin{equation}
    \textit{L}^{sjac}[\Phi] = \frac{1}{bs}\sum_{i=1}^{bs} min\bigg( \frac{(\det \nabla \Phi[\bar{x_i}] - 1)^2 }{\det \nabla \Phi[\bar{x_i}]}, \tau\bigg),
\label{eq:sjacreg}
\end{equation}

which yields an equal penalty for equal growth or shrinkage. As this penalty grows to infinity in folding vector fields, we clip the penalty to $\tau=10$. This yields a more stable objective than an asymmetric Jacobian determinant penalty at negligible additional computational cost. 

Finally, we introduce the cycle-consistency terms

\begin{equation}
\begin{split}
    \textit{L}^{cycle}_{\mathcal{F}\rightarrow\mathcal{B}} = \frac{2}{bs}\ \sum_{i=1}^{bs / 2} \  &[\PHIB(\PHIF(\vec{x}_i))-\vec{x}_i]^2 \\
    \textit{L}^{cycle}_{\mathcal{B}\rightarrow\mathcal{F}} = \frac{2}{bs}\sum_{i=bs/2}^{bs} &[\PHIF(\PHIB(\vec{x}_i))-\vec{x}_i]^2,
\end{split}
\label{eq:cycleloss}
\end{equation}

which penalize the square norm of the cycle-error vectors, encouraging $\PHIF$ and $\PHIB$ to behave like corresponding approximate inverse functions. This allows both networks to regularize each other, explicitly penalizing large spatial fluctuations in the vector fields.

Previous work on registration INRs employed a bending energy penalty\cite{rueckert1999nonrigid} as the regularization term:

\begin{equation}
\centering
\begin{split}
    &\textit{L}^{bend}[\Phi]= \frac{1}{bs} \sum_{i=1}^{bs}  \\
    \Bigg(\bigg( \frac{\partial^{2}\Phi[\bar{x_i}]}{\partial\it{x}^{2}} \bigg)^{2} &+ \bigg( \frac{\partial^{2}\Phi[\bar{x_i}]}{\partial\it{y}^{2}} \bigg)^{2} + \bigg( \frac{\partial^{2}\Phi[\bar{x_i}]}{\partial\it{z}^{2}} \bigg)^{2} + \\
    2\Bigg[ \bigg( \frac{\partial^{2}\Phi[\bar{x_i}]}{\partial\it{x}\partial\it{y}} \bigg)^{2} &+ \bigg( \frac{\partial^{2}\Phi[\bar{x_i}]}{\partial\it{x}\partial\it{z}} \bigg)^{2} + \bigg( \frac{\partial^{2}\Phi[\bar{x_i}]}{\partial\it{y}\partial\it{z}} \bigg)^{2}\Bigg]\Bigg),
\end{split}
\label{eq:bendreg}
\end{equation}

which enforces a smooth vector field by minimizing the second order derivatives of the INR around each sample. We compare two variations of our method: one where the INRs are regularized with the proposed symmetric Jacobian determinant regularization, and one where the INRs are regularized with this bending energy penalty.

\subsection{Cycle-consistent inference}\label{sec:cc_inference}
By leveraging the results from both the forward and the backward transformations, we seek to derive a more accurate forward estimate for each coordinate. The multiple-differentiable property of the \textsc{SIRENs} enables the use of Taylor expansions to derive an accurate local estimate for the inverse of the backward transformation, yielding a second estimate for the forward function. Leaning on the concept of ensembling\cite{hansen1990neural}, we evaluate a consensus of these two estimates to refine the result. This result is then given by

\begin{equation}
    \vec{x}_{result} = \frac{1}{2}\bigg( \PHIF(\vec{x}) + \PHIB^{\minus1}(\vec{x}) \bigg)
\label{eq:cycle_infer}
\end{equation}

where $\PHIB^{\minus1}(\vec{x})$ is the second-order Taylor expansion around $\PHIB(\PHIF(\vec{x}))$:

\begin{equation}\centering
\begin{split}
    \PHIB^{\minus1}(\vec{x}) =\ &\PHIB^{\minus1}[\PHIB(\PHIF(\vec{x}))] + \\
    \nabla &\PHIB^{\minus1}[\PHIB(\PHIF(\vec{x}))] \cdot (\vec{x} - \PHIB(\PHIF(\vec{x})))\ + \\
    \frac{1}{2}\nabla^{2} & \PHIB^{\minus1}[\PHIB(\PHIF(\vec{x}))] \cdot (\vec{x} - \PHIB(\PHIF(\vec{x})))^2,
\end{split}
\label{eq:taylor_expand_1}
\end{equation}

which simplifies to 

\begin{equation}
\begin{split}
    \PHIB^{\minus1}(\vec{x}) =\ &\PHIF(\vec{x})\ +\\
    \nabla^{\minus1} &\PHIB(\PHIF(\vec{x})) \cdot (\vec{x} - \PHIB(\PHIF(\vec{x})))\ + \\
    \frac{1}{2}\nabla^{\minus2} & \PHIB(\PHIF(\vec{x})) \cdot (\vec{x} - \PHIB(\PHIF(\vec{x})))^2,
\end{split}
\label{eq:taylor_expand_2}
\end{equation}

where $\nabla^{\minus1} \PHIB$ and $\nabla^{\minus2} \PHIB$ represent the inverse of the Jacobian and Hessian matrices at each point $\PHIF(\vec{x})$. The norm of vector $\PHIF(\vec{x}) - \PHIB^{\minus1}(\vec{x})$ represents the uncertainty regarding the deformation vector at coordinate $\vec{x}$. The full inference procedure is visualized in Fig.~\ref{fig:cc_inference}.

\section{Experiments and results}\label{sec:experiments}
First, we evaluate the registration accuracy and the optimization robustness of our proposed cycle-consistent INRs on the lung CT dataset. We compare the obtained results to results from INRs optimized with several non-cycle-consistent regularization terms. Afterwards, we analyse the added value of our proposed consensus-based inference method and \textcolor{darkred}{investigate the quality and robustness of our proposed uncertainty metric.} Finally, we perform additional experiments to quantify the optimization robustness in a set of abdominal 4D MRI, evaluating the generalizability of our results with respect to different data.

\subsection{Optimization details}\label{sec:experiments_optimization}
Before optimizing the INRs, image coordinates are rescaled to $[\minus1,1]$ as in~\cite{wolterink2021implicit}. Each INR contains $3$ hidden layers of $256$ neurons and is optimized for $2,500$ epochs using the Adam optimizer\cite{kingma2014adam} with a learning rate of $10^{\minus 4}$ and a batch size of $10,000$ per INR. This yields a total batch size of $20,000$ per cycle-consistent set of INRs, as samples for each INR are drawn independently from each domain. We use $\alpha=0.05$ for (symmetric) Jacobian determinant regularization and $\alpha=10$ for bending energy penalty. To select the cycle-consistency weight $\beta$, we evaluated the cycle-consistency term for a set of non-cycle-consistent INRs optimized on a sample in the lung dataset. Matching the order of magnitude to $L^{reg}_{\Phi}$ after the first epoch yielded our selected value of $\beta = 10^{\minus 3}$.

\begin{figure}[!t]
\centering
\includegraphics[width=8cm]{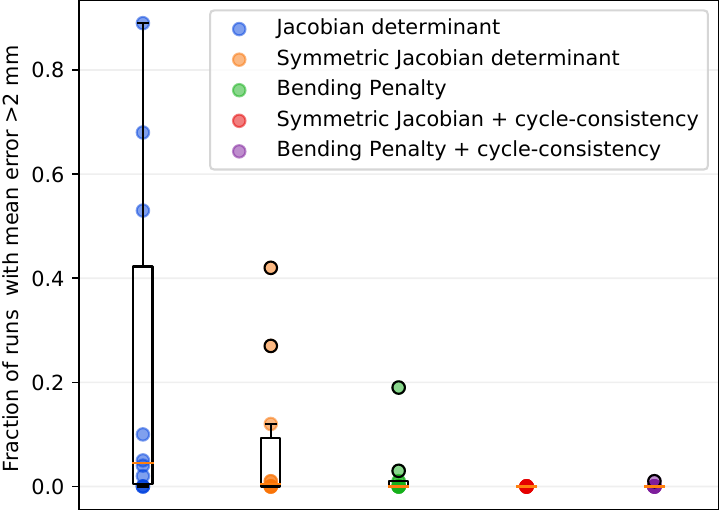}
\caption{Fraction of runs for each case in the lung CT dataset that resulted in failed registrations, estimated for both directions \textcolor{darkred}{50 times} (i.e. 50 inspiration-to-expiration results and 50 expiration-to-inspiration results \textcolor{darkred}{with different random initializations} per case for each method). Each coloured dot represents one patient. Failure is defined as a mean landmark error over 2~mm.}
\label{fig:stability_dirlab}
\end{figure}

\begin{table}[t]
\centering
\caption{Percentage of failed results in the lung CT dataset (mean landmark error over 2~mm).}
\label{tab:failed_seeds}
\begin{tabular}{lc}
\toprule
Regularization strategy & Failure rate (\%) \\
\midrule
Jacobian determinant  & 23.1  \\
Symmetric Jacobian determinant    & 8.3 \\
Bending penalty (Wolterink et al.~\cite{wolterink2021implicit})    & 2.4 \\
\midrule
Symmetric Jacobian determinant + cycle-consistency   & \bf{0.0}  \\
Bending penalty + cycle-consistency   & 0.1  \\
\bottomrule
\end{tabular}
\end{table}

\begin{figure*}[!t]
\centering
\includegraphics[width=0.87\linewidth]{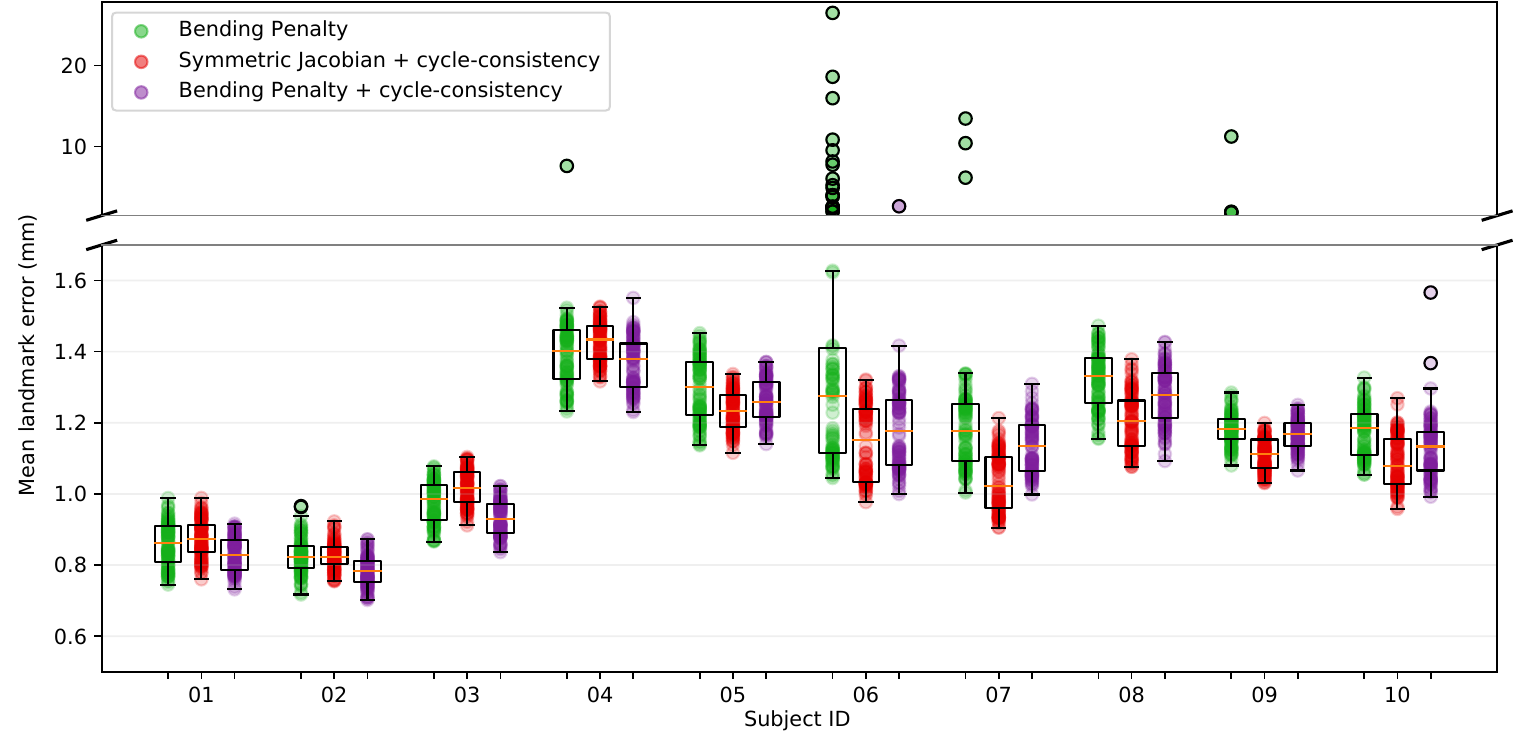}
\caption{Results on the lung CT dataset for three regularization methods. Mean landmark registration error in mm for 10 subjects, repeated with 50 random seeds, for both directions. Green: single INR with bending penalty (i.e. the method proposed in Wolterink et al.~\cite{wolterink2021implicit}), red: cycle-consistent method with symmetric Jacobian regularization, purple: cycle-consistent method with bending penalty.}
\label{fig:baseline_vs_cc}
\end{figure*}

%%%
% combined uncertainty, quali and quanti of failed seed
%%%
\begin{figure*}[b]
\centering
\includegraphics[width=0.875\linewidth]{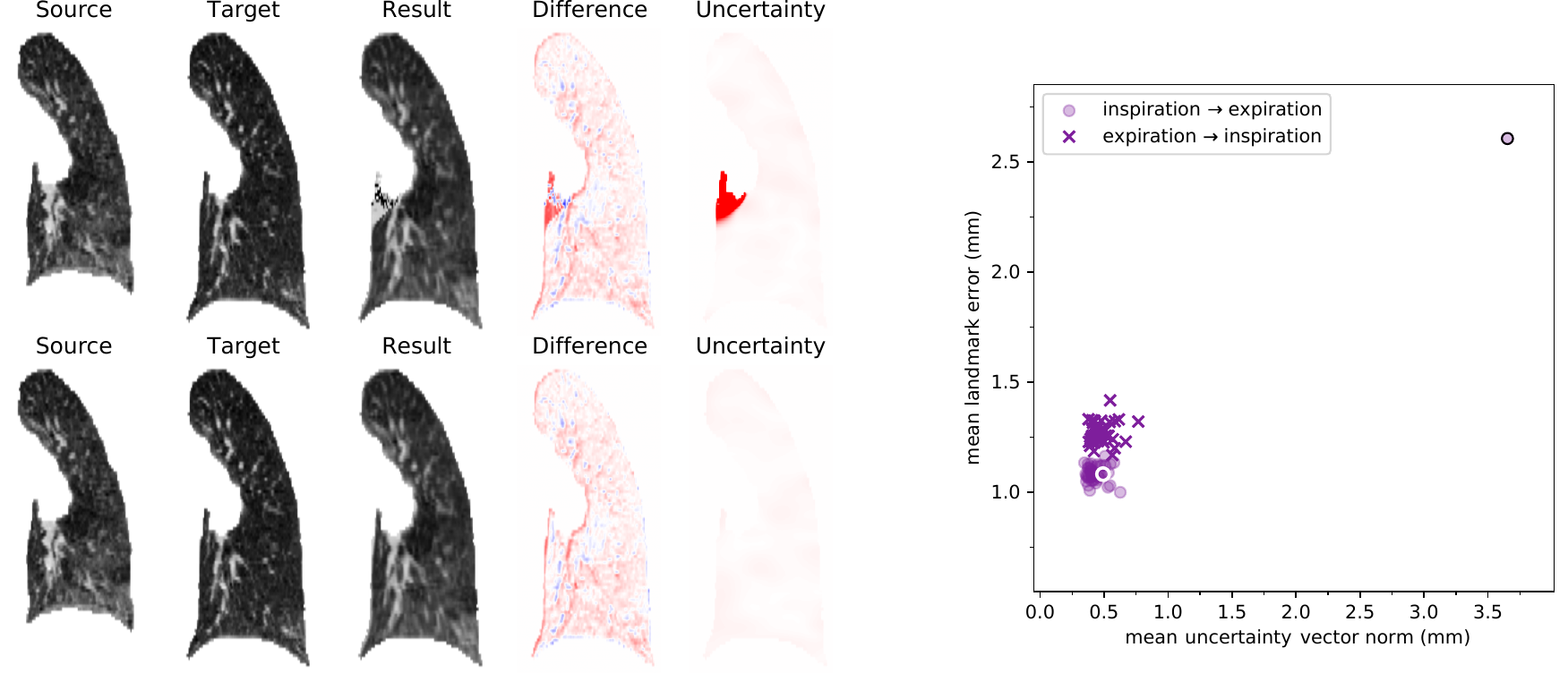}
\setcounter{figure}{5}
\caption{Left: Qualitative results in the lung CT dataset. This lung corresponds to the subject in which one of the cycle-consistent runs resulted in a failed registration. The top row shows the failed result, the bottom row shows a typical result from a different \textcolor{darkred}{run} in the same lung. The uncertainty map shows the norm of uncertainty vector $\PHIB^{\minus 1}(\bar{x})\minus\PHIF(\bar{x})$ at each coordinate within the lung mask (saturation corresponds to uncertainty, maximum saturation at 10 mm); this correctly flags the failed area near the bronchus in the failed result. Right: The mean landmark error plotted against the mean uncertainty vector norm for all runs with a bending penalty and cycle-consistency regularization for the same image, for both directions. Failed and typical results shown on the left are indicated in the scatter plot by black and white outlines, respectively.}%
\label{fig:confidence_dirlab_plus_quali}
\end{figure*}

Samples are drawn from the available foreground masks to focus the weight updates on the domains containing the image parts of interest. The foreground is defined by automatically generated lung masks from \cite{hofmanninger2020automatic} for the lung CT dataset. For the abdominal MRI dataset, the foreground is defined by body masks generated by thresholding and using morphological hole-filling. As the FOV in these scans is smaller than the imaged organ, the outer two coronal slices are excluded from the body mask in order to avoid large numerical errors near the image border. Experiments were performed on consumer hardware (Nvidia RTX 2080~Ti and a single Intel 6128 core clocked at 3.4GHz).

\subsection{Lung registration}\label{sec:experiments_dirlab}
In the lung registration dataset, we optimize 50 sets of registration INRs using different random seeds, registering the lungs in the inspiration phase to the lungs in the expiration phase and vice-versa. To quantify the registration accuracy, we evaluate the mean landmark registration error for both the expiration-to-inspiration and the inspiration-to-expiration direction. We then quantify the fraction of results that underperform due to optimization failure. To evaluate the impact of our proposed optimization method, we repeat this experiment for five different regularization strategies: asymmetric Jacobian determinant regularization (Eq.~\ref{eq:jacreg}), symmetric Jacobian determinant regularization (Eq.~\ref{eq:sjacreg}), a bending penalty (Eq.~\ref{eq:bendreg}), and combinations of the proposed cycle-consistency regularization (Eq.~\ref{eq:cycleloss}) with both the symmetric Jacobian determinant regularization and with the bending penalty, respectively. The INRs without cycle-consistent regularization are optimized separately. The setting with bending penalty, without cycle-consistency regularization is equal to the method proposed in Wolterink et al.~\cite{wolterink2021implicit}, which is the current state-of-the-art in deformable registration using INRs.

Results with an average landmark registration error of more than 2~mm are considered failed registration results; this threshold was chosen as the average distance to the adjacent voxel neighbourhood including in-plane diagonals (i.e. results are considered registration failures if the mean landmark error is larger than the average one-voxel distance). The fraction of failed results for each subject is plotted in Fig.~\ref{fig:stability_dirlab}. These results are summarized in Table~\ref{tab:failed_seeds}. The full results for the two cycle-consistent regularization strategies are shown in Fig.~\ref{fig:baseline_vs_cc}, along with the best non-cycle-consistent baseline. 

While the non-cycle-consistent bending penalty outperforms the non-cycle-consistent symmetric Jacobian determinant regularization, the cycle-consistent version of the symmetric Jacobian is the most stable strategy overall. In this setting, none of the 1,000 registration results yielded an average error over 2~mm. In the cycle-consistent setting optimized with a bending penalty, one of the registration results exceeds this threshold. The errors in this result are visualized in Fig.~\ref{fig:confidence_dirlab_plus_quali}. 

To evaluate the value of the uncertainty vector norms $|\PHIB^{\minus 1}(\bar{x})\minus\PHIF(\bar{x})|$ as a quality control metric, the right side of Fig.~\ref{fig:confidence_dirlab_plus_quali} shows the mean landmark error for all results from cycle-consistent INRs optimized with a bending penalty against the mean uncertainty vector norm for the image where the failed result occurred. This shows the outlier is linearly separable from the other 99 results, with a large margin for the threshold.

An interesting phenomenon can be observed for the failed cycle-consistent result. Two INRs were optimized for this random seed (i.e. both $\PHIF$ and $\PHIB$), yet only the inspiration-to-expiration result exhibits a large mean landmark error. 
These errors are caused by a small irregular region in $\PHIB$ in this seed, resulting in exploding gradients when this function is inverted. This effect does not appear in the opposite transformation, where $\PHIF$ is inverted instead of $\PHIB$.

\begin{table}[t]
\centering
\caption{\textcolor{darkred}{Mean landmark error comparison with other methods for inspiration-to-expiration registration.}}
\label{tab:compare_lmerr}
\begin{tabular}{lc}
\toprule
\textcolor{darkred}{Method} & \textcolor{darkred}{mean $\pm$ std (mm)} \\
\midrule
\textcolor{darkred}{DLIR\cite{de2019deep}}                               &\textcolor{darkred}{$2.64 \pm 4.32$}\\
\textcolor{darkred}{VoxelMorph~\cite{balakrishnan2019voxelmorph}}        &\textcolor{darkred}{$2.26 \pm 2.30$} \\
\textcolor{darkred}{CycleMorph~\cite{kim2021cyclemorph}}                 &\textcolor{darkred}{$2.19 \pm 2.26$} \\
\textcolor{darkred}{CNN with anatomical constraints\cite{hering2021cnn}} &\textcolor{darkred}{ $1.14 \pm 0.76$} \\
\textcolor{darkred}{Uniform B-Splines~\cite{berendsen2014registration}}  &\textcolor{darkred}{ $1.36 \pm 1.01$} \\
\textcolor{darkred}{CorrField~\cite{heinrich2015estimating}}             &\textcolor{darkred}{ $1.12 \pm 1.08$} \\
\textcolor{darkred}{Keypoint correspondence optimization~\cite{ruhaak2017estimation}} &\textcolor{darkred}{$0.94 \pm 1.06$} \\
\midrule
 \textcolor{xkcdorange}{$\bullet$} \textcolor{darkred}{INR + Symmetric Jacobian det.}            &\textcolor{darkred}{ $1.27 \pm 2.27$} \\
 \textcolor{xkcdgreen}{$\bullet$} \textcolor{darkred}{INR + Bending penalty} &\textcolor{darkred}{$1.10 \pm 1.42$} \\
 \textcolor{xkcdred}{$\bullet$} \textcolor{darkred}{INR + Symmetric Jacobian det. + cycle} &\textcolor{darkred}{$1.04 \pm 1.11$} \\
 \textcolor{xkcdpurple}{$\bullet$} \textcolor{darkred}{INR + Bending penalty + cycle} &\textcolor{darkred}{$1.06 \pm 1.34$} \\
\bottomrule
\end{tabular}
\end{table}

\subsection{Registration performance comparison}\label{sec:performance_comparison}
\textcolor{darkred}{We compare the registration performance of the presented method with a number of existing methods, both learning-based and optimization-based, with results on the publicly available DIR-Lab benchmark. The mean landmark error results for inspiration-to-expiration registration are shown in Table~\ref{tab:compare_lmerr}. For DLIR\cite{de2019deep}, Uniform B-Splines~\cite{berendsen2014registration},  anatomically constrained CNNs~\cite{hering2021cnn} and keypoint correspondence optimization~\cite{ruhaak2017estimation}, we show the results reported in their respective papers. Results for CorrField~\cite{heinrich2015estimating}, VoxelMorph\cite{balakrishnan2019voxelmorph} and CycleMorph\cite{kim2021cyclemorph} were generated using their respective publicly available implementations, as these works did not report results on the original DIR-Lab set. The latter two methods were trained 10 times using cross-validation, where 8 cases were used in training, one in validation and one in testing. As these methods do not support incorporating foreground masks, all cases were cropped to a rough bounding-box of the lungs to reduce the required amount of GPU memory and intensities were clipped to a window as in \cite{de2019deep}. Models were trained on each possible pair of timepoints (i.e. 800 training pairs per fold), for 100 epochs on an Nvidia RTX 3090. The test image in each fold was evaluated on the epoch with the lowest landmark error in validation.}
\textcolor{darkred}{For the INR-based experiments, errors are averaged over all 50 runs.} \textcolor{darkgreen}{These results are discussed in section~\ref{sec:discussion}.}

\subsection{Sensitivity to alternative design choices}\label{sec:hyper_ablation}
\textcolor{darkred}{We evaluate the sensitivity of the presented method to the selected hyper-parameters and design decisions in terms of the total registration error (TRE) and the failure rate (defined as the fraction of runs that result in a mean registration error over 2 mm, as in section \ref{sec:experiments_dirlab}). Table~\ref{tab:hyperablation} shows the TRE on the DIRLAB dataset (aggregated for both inspiration-to-expiration and expiration-to-inspiration registration, median across 50 runs) and the failure rate, for several variations of Jacobian determinant factor $\alpha$ and cycle-consistency factor $\beta$, as well as for registration without lung masks and registration with ReLU-based networks rather than SIRENs. For the cycle-consistent inference of the ReLU-based networks, a first-order Taylor expansion was used as these networks are not differentiable multiple times. For the experiment without masks, images were clipped to the window proposed in \cite{de2019deep} and coordinates were sampled uniformly from a rough bounding box around the lungs.}

\textcolor{darkred}{The median TRE denotes the typical performance and failure rate indicates the reliability of the method. The results show that the proposed method is not very sensitive to hyper-parameter changes, with all variations resulting in better performance and reliability than the non-cycle-consistent baseline.}

\begin{table}[t]
\centering
\caption{\textcolor{darkred}{Median registration error and failure rate with alternative settings for symmetric Jacobian determinant regularization.}}
\label{tab:hyperablation}
\begin{tabular}{ccc}
\toprule
\textcolor{darkred}{Setting} & \textcolor{darkred}{TRE (mm)} & \textcolor{darkred}{Failure rate (\%)}\\
\midrule
 \textcolor{darkred}{Single INR ($\alpha = 0.05$, $\beta = 0$)} & \textcolor{darkred}{$1.14$} & \textcolor{darkred}{$8.3$} \\
 \textcolor{darkred}{Proposed ($\alpha = 0.05$, $\beta = 1e\minus3$)} & \textcolor{darkred}{$1.07$} & \textcolor{darkred}{$0$} \\
 \midrule
 \textcolor{darkred}{Jacobian det. $\alpha = 0.50$} & \textcolor{darkred}{$1.11$} & \textcolor{darkred}{$0$} \\
 \textcolor{darkred}{Jacobian det. $\alpha = 0.10$} & \textcolor{darkred}{$1.08$} & \textcolor{darkred}{$0$} \\
 \textcolor{darkred}{Jacobian det. $\alpha = 0.025$} & \textcolor{darkred}{$1.07$} & \textcolor{darkred}{$0.3$} \\
 \textcolor{darkred}{Jacobian det. $\alpha = 0.005$} & \textcolor{darkred}{$1.07$} & \textcolor{darkred}{$1.8$} \\
 \midrule
 \textcolor{darkred}{cycle $\beta = 1e\minus 2$} & \textcolor{darkred}{$1.08$} & \textcolor{darkred}{$0.4$} \\
 \textcolor{darkred}{cycle $\beta = 2e\minus 3$} & \textcolor{darkred}{$1.07$} & \textcolor{darkred}{$0$} \\
 \textcolor{darkred}{cycle $\beta = 5e\minus 4$} & \textcolor{darkred}{$1.07$} & \textcolor{darkred}{$0$} \\
 \textcolor{darkred}{cycle $\beta = 1e\minus 4$} & \textcolor{darkred}{$1.06$} & \textcolor{darkred}{$0.4$} \\
 \midrule
 \textcolor{darkred}{ReLU network}       & \textcolor{darkred}{$1.22$} & \textcolor{darkred}{$0$} \\
 \textcolor{darkred}{No masks}       & \textcolor{darkred}{$1.11$} & \textcolor{darkred}{$0$} \\

\bottomrule
\end{tabular}
\end{table}

\subsection{Small intestine registration}\label{sec:experiments_motac}
We use the proposed method to propagate intestinal centerline annotations across multiple timepoints in the abdominal 4D MRI dataset; this is a particularly challenging task, as none of the points on the centerlines can be matched to specific landmarks or local image features. In the setting of registering larger numbers of timepoints, we consider optimizing INRs with a bending penalty clinically unfeasible: this takes approximately 12 minutes per set on current consumer hardware, as opposed to 90 seconds for optimizing INR sets with (symmetric) Jacobian determinant regularization. Hence, we only compare symmetric Jacobian determinant regularization with and without cycle-consistent regularization in this setting.

For each image series in the dataset, we optimize sets of registration INRs mapping between $t_0$ and each timepoint during breath-hold, repeated with 50 different random seeds. Due to the strong anisotropy of the field-of-view in these images, we use zero padding in the smallest dimension before rescaling the coordinates to $[\minus1,1]$, yielding an isotropic coordinate space.

\begin{figure}[!t]
\centering
\includegraphics[width=0.92\linewidth]{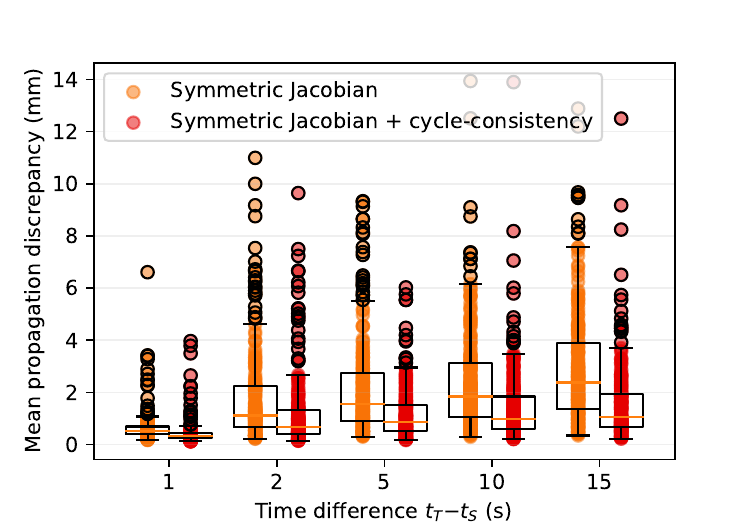}
\caption{A comparison of the mean propagation discrepancy across runs with different random seeds for each intestinal centerline segment in the abdominal MRI dataset, for increasing intervals between registered timepoints.}
\label{fig:motac_mediandists}
\end{figure}

We use the resulting implicit functions to propagate all centerline segments across all timepoints. To evaluate the robustness of the method, we construct a propagation consensus for each centerline segment as the median result from all 50 seeds and we evaluate the propagation discrepancy for each centerline segment. This is defined as the median marker-to-marker distance from the results in each seed to the propagation consensus, averaged along each segment. A higher propagation discrepancy indicates more disagreement between different random seeds. The distributions of these results are shown in Fig.~\ref{fig:motac_mediandists} and summarized in Table~\ref{tab:propagation_discrepancies}. Both for small and for large time intervals, cycle-consistent optimization results in substantially lower propagation discrepancy.

\begingroup
\setlength{\tabcolsep}{5pt}
\begin{table}[b]
\centering
\caption{Centerline propagation discrepancy in the Cine-MRI dataset for several time intervals, for both single INR and the proposed cycle-consistent optimization (lower is better).}
\label{tab:propagation_discrepancies}
\begin{tabular}{lcccc}
\toprule
 & \multicolumn{4}{c}{Discrepancy (mm) $\pm$ stdev} \\
Strategy & 1s & 5s & 10s & 15s \\
\midrule
Single INR  & $0.68 \pm 0.68$   & $2.26 \pm 2.05$  & $2.47 \pm 2.05$    & $2.99 \pm 2.31$ \\
Proposed    & $0.45 \pm 0.54$   & $1.21 \pm 1.06$  & $1.48 \pm 1.54$    & $1.62 \pm 1.57$ \\
\bottomrule
\end{tabular}
\end{table}
\endgroup

\begin{figure}[!t]
\centering
\includegraphics[width=0.8\linewidth]{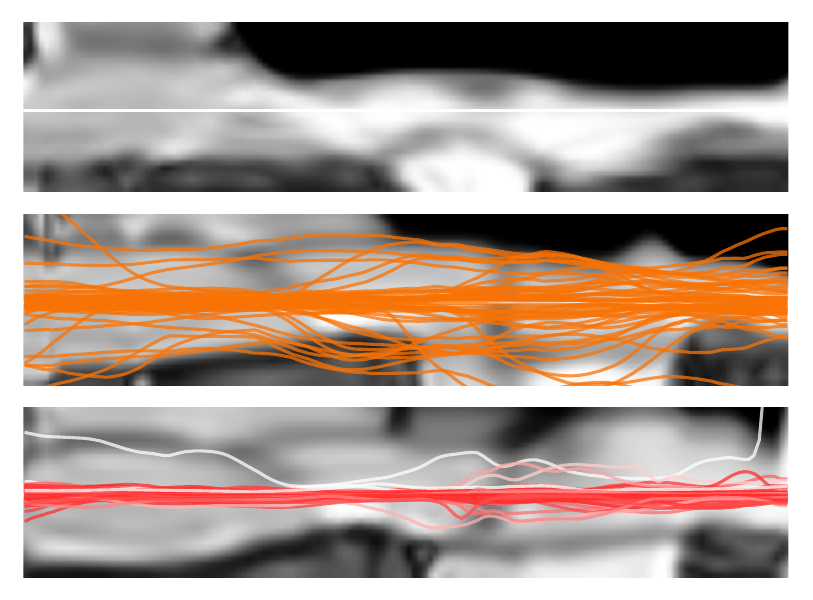}
\caption{Untangled representations of a typical small intestine segment. Top: the segment at $t_0$, untangled around the manual centerline annotation. Middle: the same segment at $t_{+15s}$, untangled around the non-cycle-consistent propagation consensus. Results for 50 seeds projected in orange. Bottom: the segment at $t_{+15s}$, untangled around the cycle-consistent propagation consensus. Results for 50 cycle-consistent INR sets, saturation dependent on mean uncertainty vector norm along the segment (zero saturation for mean uncertainty vector norms over 5mm).}
\label{fig:mpr_result}
\end{figure}

\begin{figure}[t]
\centering
\includegraphics[width=0.9\linewidth]{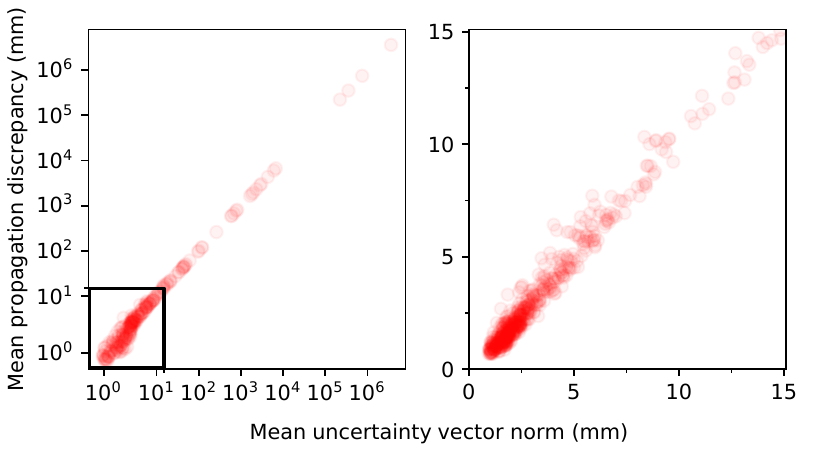}
\caption{The mean propagation discrepancy (i.e. distance to the propagation consensus) plotted against the proposed uncertainty metric when registering timepoints with an interval of 15 seconds. Each data point represents the mean results from one cycle-consistent INR pair. Left: log plot showing all results. Right: zoomed-in linear visualisation of the area indicated by a black rectangle in the left plot.}
\label{fig:confidence_motac}
\end{figure}

An example of the results is shown in Fig.~\ref{fig:mpr_result}, which shows an untangled representation of one of the intestinal segments with projections of propagated centerlines for INRs optimized both with and without cycle-consistent regularization. The cycle-consistent results both contain fewer failed results, and the successful results form a tighter bundle, indicating a more reproducible result. Additionally, the diverging seeds have large mean uncertainty vector norms, indicated by low saturation in the figure, meaning the failures that occur here are automatically detectable.

Fig.~\ref{fig:confidence_motac} visualizes the correlation between the proposed uncertainty metric and the propagation discrepancy for each centerline segment for each seed for a time interval of 15 seconds. There is a very strong linear correlation between the proposed confidence metric and the mean distance to the propagation consensus. This figure also shows a number of severe outliers, visualized in the log-plot on the left side. As in the experiments on the lung CT dataset, these errors are caused by a irregular regions in $\PHIB$, resulting in exploding gradients during function inversion.

\section{Discussion}\label{sec:discussion}
In this work, we have presented a method for deformable image registration using pairs of INRs. The method employs cycle-consistency to regularize the optimization process and to improve the accuracy of the predicted vector fields. Our experiments reveal three important properties of the presented method. The first is improved optimization stability compared to a single-INR registration method: in our experiments on the lung CT dataset, cycle-consistent optimization reduced the failure rate from 8.3\% to 0.0\% when using symmetric Jacobian determinant regularization and reduced failure rate from 2.4\% to 0.1\% when using a bending penalty. This is a clear improvement over the current state-of-the-art in deformable registration using INRs. The second property is a reliable uncertainty metric. The proposed uncertainty metric detected all large landmark errors in the experiments on the lung CT dataset, and the metric has a strong correlation with the propagation discrepancies in the abdominal MRI task. Beside being reliable, the uncertainty metric is also easily interpretable by an end-user, as it is expressed as an estimated error in the unit of the optimized vector field. Finally, our method achieves clear improvement in registration accuracy, yielding lower average landmark errors for all subjects in the lung CT dataset and lower propagation discrepancies for all time intervals for all subjects in the abdominal MRI set. \textcolor{darkred}{Furthermore, our ablation experiments show that the presented method is not very sensitive to suboptimal hyper-parameter settings, which makes it easier to use in practice.}

When optimized with a bending penalty, the cycle-consistent method yielded a single failed cycle-consistent result in the lung CT dataset, as illustrated in Fig.~\ref{fig:confidence_dirlab_plus_quali}. The landmarks that exhibit large errors in this result are near the border of the lung mask, at the connection of the left bronchus. The absence of optimization samples in surrounding voxels likely contributed to the lack of successful regularization of $\PHIB$ in this region, which resulted in exploding gradients during function inversion. 
While this behavior could be construed as a weakness of the method, we do not consider this effect problematic: the sensitivity of the inference method to irregular regions in the optimized functions ensures the uncertainty vector norm is large in these areas, enabling the detection of any optimization runs where a subset of the INR domains failed to converge to a smooth function. 

Our experiments on the lung CT dataset imply that cycle-consistent INRs optimized with symmetric Jacobian determinant regularization are slightly more stable than INRs optimized with the more computationally expensive bending energy penalty, despite the opposite being true for non-cycle-consistent INRs. Furthermore, the mean performance improvement of the cycle-consistent inference procedure compared to forward-only inference was significantly higher for INRs optimized with symmetric Jacobian determinant regularization. These results imply that symmetric Jacobian determinant regularization results in more consistently smooth functions than a bending energy penalty when either is combined with cycle-consistent regularization. 

The mean landmark error for the inspiration-to-expiration registration task is slightly lower than the landmark error in the reverse direction, as shown in Fig.~\ref{fig:confidence_dirlab_plus_quali}. This is consistently true in all of our experiments, regardless of regularization strategy. While our proposed regularization strategy is fully symmetric with respect to shrinkage and expansion, the task itself is not: an equal relative error for positioning landmarks in both phases results in a higher absolute error in the inspiration phase, as the larger lung volumes in this phase result in larger distances between relative positions. 

\textcolor{darkred}{In this data-limited setting, the proposed method substantially outperforms learning-based methods. Although the method is outperformed by dedicated lung-registration methods that incorporate domain specific knowledge in their optimization strategy~\cite{ruhaak2017estimation}, the cycle-consistent INR-based method yields more accurate registrations than the general-purpose methods that we evaluated.}

In the experiments on the abdominal MRI dataset, we observe that our method results in a large improvement in reproducibility across random seeds, reducing the variation between registration results obtained with different random initializations. It should be noted that due to the lack of ground truth in multiple timepoints, we evaluated the propagation discrepancy (i.e. the distance to the consensus). While these experiments prove improved reproducibility, they do not strictly prove improved accuracy, as the propagation discrepancy is merely an estimate of the true error. Nevertheless, from visual evaluation these estimates seem accurate, as shown in Fig.~\ref{fig:mpr_result}.

Unlike in the lung CT dataset, the cycle-consistent INRs optimized on the abdominal MRI task yield a fairly large number of outliers. The reason for this is the difficulty of the propagation task. One issue is that the static FOV of each timepoint does not perfectly match the FOV of later timepoints, resulting in an undefined propagation function at larger time intervals for any markers near the border of the FOV. Furthermore, motile activity causes changes to the topological structure of the intestines: as the intestine contracts, the visually distinct intestinal content is segmented into multiple disjoint volumes, broken up by contracting intestinal wall. This makes registration near these areas particularly challenging. Additionally, due to the difficulty of annotating intestinal centerlines, imperfections in the centerline annotations result in markers being placed outside of the center of the small intestine. Due to the rotationally symmetric appearance of the small intestine, these locations are visually ambiguous with coordinates at similar relative distances from the true centerline. This hampers reproducibility of the centerline propagation.

The proposed method uses a second-order Taylor expansion to achieve a highly accurate inversion of the backward INR. This is only possible when using networks that are differentiable multiple times, such as the SIREN network used in our experiments. However, the proposed inference method \textcolor{darkred}{can be adapted to work with ReLU models by using the first-order Taylor expansion instead, which was done for the ablation experiment in Section~\ref{sec:hyper_ablation}. This results in slightly less accurate inverse functions}, especially considering the piecewise linearity of function estimates produced by ReLU networks, but the basic principle remains the same.

\textcolor{darkred}{While the used network design results in adequate performance for the evaluated datasets, this choice is not guaranteed to be an optimal or suitable design for all registration problems. When applying this method to other datasets, adjusting the number of features and/or layers in each network to the specific problem is likely to be beneficial.}

\textcolor{darkred}{The runtime of the proposed cycle-consistent method is approximately 90 seconds on consumer hardware: 1.8x longer than registration with a single INR, which itself is quite slow compared to methods like corrField~\cite{heinrich2015estimating} with runtimes of several seconds. One of the reasons is that the optimization procedure used in this work is somewhat naive: every INR is optimized for $2,500$ epochs with a fixed learning rate, starting from a completely random initialisation and without an early stopping criterion. An interesting avenue for future work would be to improve the runtime of this procedure, for example by using meta-learning to find more favorable initializations~\cite{baum2022meta} or by improving the optimization schedule.}

Our proposed cycle-consistent regularization method is independent of the data loss metric. While we used the normalized cross correlation in our experiments, this is only applicable for unimodal registration systems. For multi-modal registration, this metric should be substituted with a modality-agnostic similarity metric such as the normalized mutual information~\cite{de2020mutual}. 

\section{Conclusion}
This work has presented a deformable registration method that leverages cycle-consistency in sets of Implicit Neural Representations that represent opposing deformation functions. Compared to using a single INR, our method yields more accurate registration results and improved reproducibility. Additionally, the method yields \textcolor{darkred}{a robust and highly interpretable confidence metric} that can be used for automatic quality control. These properties are vital for enabling the use of deformable registration methods in autonomous settings.

\bibliographystyle{IEEEtran}
\bibliography{tmi}

\end{document}